\documentclass[aps,prd,twocolumn,groupedaddress,showpacs,preprintnumbers]{revtex4}

\usepackage{amssymb}
\usepackage{graphicx}
\usepackage{epsfig}
\newcommand{\beq}{\begin{equation}} 
\newcommand{\eeq}{\end{equation}}   
\newcommand{\bea}{\begin{eqnarray}} 
\newcommand{\eea}{\end{eqnarray}}
\newcommand{\ta}[1]{#1\hspace{-.62em}/\hspace{-.07em}} 
\def\Li2{\hbox{Li}_2}

\begin{document}


\preprint{\bf TTP07-03}

\title{ Spin  asymmetries and correlations in ${\bf \Lambda}$--pair production
  \\ through the radiative return method}

\thanks{Work 
supported in part by BMBF under grant number 05HT6VKA,
EU 6th Framework Program under contract MRTN-CT-2006-035482 
(FLAVIAnet), TARI project RII3-CT-2004-506078,
Polish State Committee for Scientific Research (KBN)
under contract 1 P03B 003 28}


\author{Henryk Czy\.z}
\affiliation{Institute of Physics, University of Silesia,
PL-40007 Katowice, Poland.}

\author{Agnieszka Grzeli{\'n}ska}
\author{Johann H. K\"uhn}

\affiliation{Institut f\"ur Theoretische Teilchenphysik,
Universit\"at Karlsruhe, D-76128 Karlsruhe, Germany.}

\date{\today}

\begin{abstract}
 Exclusive baryon pair production through the radiative return
 is a unique tool to determine the electric and the magnetic form factors
 in the timelike region over a wide range of energies. The decay of 
 $\Lambda$-baryons into $p +\pi^-$ can be used to measure
 their spin and to analyze
 spin-spin correlations in the  $\bar\Lambda\Lambda$ system.
 We evaluate the spin dependent hadronic tensor, which describes this reaction,
 calculate the corresponding cross section for 
 $e^+e^- \to \Lambda\bar\Lambda$ and $e^+e^- \to \Lambda\bar\Lambda\gamma$
 with subsequent decays of  $\Lambda$ and $\bar\Lambda$ and present a compact
 formula for the correlated distribution. We demonstrate that the relative
 phase between the electric and the magnetic form factor can be determined
 in this analysis by considering single spin asymmetries and correlated
 decay distributions. The reaction is implemented in the event generator
  PHOKHARA. It is demonstrated that the relative phase can be measured 
with the event sample presently accumulated at $B$- factories.
\end{abstract}


\maketitle


\section{\label{sec1}Introduction}
Measurements of the proton--antiproton production through
 the radiative return,
 as suggested in \cite{Nowak}, have lead to a determination of the proton
 electric and magnetic form factors, $G_E$ and $G_M$, in the time-like region
 over an energy which ranges from threshold to nearly 3~GeV
 \cite{Aubert:2005cb}. The study of the angular distributions allows to
 separate the contributions from $|G_M|^2$ and $|G_E|^2$ to the cross 
 section. In the absence of longitudinally polarized beams and without
 measuring of the polarization of the produced baryons, the relative phase
 between  $G_M$ and $G_E$ does not enter the predictions \cite{Nowak}.
 In this paper we demonstrate that $\Lambda$$\bar \Lambda$-- production 
 through the radiative return can be used to fully determine modulus 
 and relative phase $\Delta\phi$
 of $\Lambda$ electric and magnetic form factors.
 We present a compact form for the hadronic tensor
 \bea
   H_{\mu\nu} = \ \langle \bar \Lambda\Lambda|{\cal J_\mu}|0\rangle \ 
               \langle 0| {\cal J}^{\dagger}_\nu | \bar \Lambda
  \Lambda \rangle\ = \ 
       J_\mu J^{\dagger}_\nu \ ,
 \eea
including its spin dependent part and analyze single--spin asymmetries
 and spin--spin correlations for $\Lambda$$\bar\Lambda$ production
through $e^+e^-$ annihilation and through the radiative return.
The self analyzing decays  $\Lambda\to p \pi^-$ and $\Lambda\to n\pi^0$
can be used to directly translate the spin asymmetries and correlations
into predictions for the decay products. The spin component normal
 to the production plane is a measure of $\sin(\Delta\phi)$, while
 spin--spin correlations can be used to resolve the remaining twofold
 ambiguity and determine the relative phase completely.

 In addition to this theoretical study we construct an event generator
 (an extension of the event generator PHOKHARA \cite{Rodrigo:2001jr,Rodrigo:2001kf,Kuhn:2002xg,Czyz:PH03,Nowak,Czyz:PH04,Czyz:2004nq,Czyz:2005as}),
 which includes the spin correlated decays of $\Lambda$ and $\bar\Lambda$.
 We adopt a simple model for the form factor and demonstrate, that the 
 determination of the phase is indeed feasible. Once required by an experiment,
 the approach can be trivially extended to the full baryon octet.

\section{\label{sec2} The method}

  The reactions
 $e^+(p_1)e^-(p_2)\to \bar\Lambda(q_1,S_1) \Lambda(q_2,S_2)$ 
 and $e^+(p_1)e^-(p_2)\to \bar\Lambda(q_1,S_1) \Lambda(q_2,S_2) \gamma(k)$
 (with photon emitted from one of the initial states)
 are both in 
 Born approximation fully described by electromagnetic
 current with two independent form factors
 \begin{eqnarray}
&&\kern-30pt J_\mu = - ie \cdot \bar{u}(q_2,S_2) \nonumber \\
&&\times \biggl( F_1^\Lambda(Q^2) \gamma_\mu
- \frac{F_2^\Lambda(Q^2)}{4 m_\Lambda}[\gamma_\mu,\ta{Q} \  ]   \biggr)  
 \ \ v(q_1,S_1), 
\label{current}
 \end{eqnarray}
related to the electric $G_E$ and magnetic $G_M$ form factors by
 \begin{eqnarray}
G_M = F_1^\Lambda + F_2^\Lambda, \ 
G_E = F_1^\Lambda + \tau F_2^\Lambda, \\ \ {\rm with} \ 
 \tau =\frac{Q^2}{4m_\Lambda^2}\  {\rm and}\  Q = q_1 + q_2.\nonumber
\label{GEGM}
 \end{eqnarray}
The $q_i,S_i, \ \ i=1,2 $ are the four momenta and the spin four vectors
 of $\bar\Lambda$ and $\Lambda$ with the properties  $S_i.q_i = 0, \ S_i^2=-1$.
 We will not consider the
 initial lepton polarizations and denote their four momenta by
 $p_1$ (positron)
 and $p_2$ (electron). 
 The resulting squared amplitudes, 
 which describe the two processes under consideration, 
  are then given by
   \bea
    |M^{0,1}|^2 =  L^{0,1}_{\mu\nu}H^{\mu\nu} \ ,
\label{amp}
   \eea
  where $0$($1$) stands for the process without (with) photon.
The hadronic tensor, which is identical for both cases, and defined as 
   \bea
  H_{\mu\nu} = J_\mu J^{\dagger}_\nu =  H_{\mu\nu}^S +H_{\mu\nu}^A \ ,
\label{hadtens}
   \eea

is naturally splitted into symmetric ($H_{\mu\nu}^S$) 
 and antisymmetric ($H_{\mu\nu}^A$) parts
  defined as
   \bea
  H_{\mu\nu}^S = \frac{1}{2} \left(J_\mu J_\nu^{\dagger}
     + J^{\dagger}_\mu J_{\nu} \right)  , \ 
 H_{\mu\nu}^A = \frac{1}{2} \left(J_\mu J^{\dagger}_\nu
     - J^{\dagger}_\mu J_{\nu} \right) \ .
\label{tensdefas}
   \eea 
They are equal to
\begin{widetext}
 \begin{eqnarray}
&\kern-10pt\frac{1}{e^2}H^S_{\mu \nu}
  = \ \frac{|G_M|^2}{2} \biggl(  Q.S_2 \  \{ Q_\mu, S_{1 \nu} \}_+ 
+  Q.S_1 \  \{ Q_\mu , S_{2 \nu} \}_+  
-   S_1 . S_2  \ Q_\mu \ Q_\nu 
- Q^2 \ \{ S_{1 \mu} , S_{2 \nu} \}_+  \biggr)
\nonumber \\
&+  |G_M|^2 ( \frac{Q^2}{2} \ S_1.S_2   -  Q.S_1 \ Q.S_2 )\ \ g_{\mu \nu}
+ \biggl(  \frac{|G_M - G_E|^2}{m_\Lambda^2 (\tau - 1)^2}  \  Q.S_1 \ Q.S_2 \ 
+ \frac{ 2\tau}{\tau - 1}\biggl(|G_M|^2 - \frac{1}{\tau} |G_E|^2 \biggr) \ S_1 . S_2  \biggr) \ \ q_\mu \ q_\nu 
\nonumber \\
&+  \frac{ \tau  |G_M|^2 - Re(G_M G_E^*) }{\tau - 1}
( Q.S_2 \ \ \{q_\mu, S_{1 \nu}\}_+ -  Q.S_1 \ \ \{q_\mu,S_{2 \nu}\}_+)
+ \ \frac{Im(G_M G_E^*)}{m_\Lambda (\tau-1) }
  \ \{\varepsilon_{\beta \gamma \delta \mu }\ Q^{\beta} q^{\gamma}
    \ (S_{1}^{ \delta} + S_{2}^{ \delta}),q_{\nu}\}_+
+ \frac{1}{4}H^U_{\mu\nu} \ ,\nonumber \\&
\label{hadtensS}
 \end{eqnarray}
 \begin{eqnarray}
&\frac{1}{e^2}H^A_{\mu \nu} =
  i \frac{  Im(G_M G_E^*) }{\tau - 1}
( Q.S_2 \ \ \{q_\mu ,\ S_{1 \nu}\}_- -  Q.S_1 \ \ \{q_\mu, \ S_{2 \nu}\}_-)
\nonumber \\ &
- \ i \ \frac{Re(G_M G_E^*)}{m_\Lambda (\tau-1) } \
  \{ \varepsilon_{\beta \gamma \delta \mu }
 \ Q^{\beta} q^{\gamma}  \ (S_{1}^{ \delta} + S_{2}^{ \delta} ),q_{\nu}\}_-
+ \ i \ \frac{|G_M|^2}{ 2m_\Lambda (\tau - 1)} \varepsilon_{\beta \gamma \mu \nu } 
\ Q^{\beta} q^{\gamma}  \ ( Q.S_2 - Q.S_1 ) \ ,
\label{hadtensA}
 \end{eqnarray}
\end{widetext}
where $q = (q_2 - q_1)/2$, 
 $\{a_\mu,b_\nu\}_\pm\equiv a_\mu b_\nu\pm b_\mu a_\nu$ and
\bea
&H^U_{\mu\nu} = 2 |G_M|^2 ( Q_\mu Q_\nu - g_{\mu \nu} Q^2)\nonumber \\
  &- \frac{ 8\tau}{\tau - 1}\biggl(|G_M|^2 - \frac{1}{\tau} |G_E|^2 \biggr)
  q_\mu q_\nu 
\label{hadtensunpol}
\eea
 is the unpolarized hadronic tensor.

 In the first step, for illustration and further reference, we shall discuss
 baryon production in electron--positron annihilation without photon radiation.
 The application to the radiative return will be presented subsequently.

 The lowest order leptonic tensor,
 when one  neglects the electron mass, is equal to
\bea
 L^{0}_{\mu\nu} &=& \frac{\pi\alpha}{s^2} 
\left(- 2 s g^{\mu\nu} + 4 \{p_1^{\mu},p_2^{\nu}\}_+\right) \nonumber\\
 &=& \frac{2\pi\alpha}{s} \ \ {\rm diag}(0,1,1,0) 
\ ,
\label{l0}
\eea
 while the explicit form of $ L^{1}_{\mu\nu}$ 
 (describing the radiative return),
 together with definitions and
 notation, can be found for instance
 in \cite{Kuhn:2002xg,Czyz:2000wh}.
 In both cases the leptonic tensor is symmetric 
 (apart from small imaginary part coming from the radiative corrections),
  hence the asymmetric part of the hadronic tensor
 does not contribute to the cross section
\begin{equation}
d\sigma(e^+e^-\to\bar\Lambda\Lambda)
  = \frac{1}{2s}L^0_{\mu\nu}H^{\mu\nu}d\Phi_2(p_1+p_2;q_1,q_2) \ ,
\label{sigma0}
\end{equation}
 which can be written in the following compact form

\begin{widetext}
  \begin{eqnarray}
&\kern-30ptL^0_{\mu\nu} H^{\mu\nu} = \frac{4\pi^2\alpha^2}{s^2}\Biggl\{
  \ 16 \ q.p \ \frac{\tau |G_M|^2 - Re(G_M G_E^*)}{\tau-1} \ 
( Q.S_1 \ p.S_2 - Q.S_2 \ p.S_1 )+ 2 \ s \ |G_M|^2 ( s + 4 \ p.S_1 \ p.S_2) 
 \nonumber
\\ 
&
-\biggl[  \frac{ 2\tau s}{\tau - 1}\biggl(|G_M|^2 - \frac{1}{\tau} |G_E|^2 \biggr) + \ \frac{8 |G_M-G_E|^2}{m_\Lambda^2 (\tau-1)^2} 
\ (q.p)^2 \biggr] \ Q.S_1 \ Q.S_2
\nonumber \\ 
& - \frac{ 2\tau }{\tau - 1}\biggl(|G_M|^2 - \frac{1}{\tau} |G_E|^2 \biggr)
\ ( 8 \ (q.p)^2 + 2 \ s \ m_\Lambda^2 - \frac{s^2}{2} ) (S_1.S_2 - 1) 
 -   \ q.p \ \frac{16 \ Im(G_M G_E^*)}{m_\Lambda (\tau-1)} \ 
\varepsilon_{\beta \gamma \mu \nu } \ Q^{\beta}  q^{\gamma}  
(S_1^{\mu} + S_2^{\mu}) p^{\nu} \Biggr\}, \ \nonumber \\ &
\label{tensors} 
\end{eqnarray}
\end{widetext}
where $ p = (p_2 - p_1)/2$ and $s = (p_1+ p_2)^2=Q^2$.
This result becomes more transparent, if we use the spin vectors
 of the $\Lambda$ and $\bar\Lambda$ in their respective rest frame

\begin{widetext}
  \bea
&\kern-80pt L^0_{\mu\nu} H^{\mu\nu} =4\pi^2\alpha^2\Biggl\{
 |G_M|^2\left(1+\cos^2\theta_{\bar\Lambda}\right)
 +\frac{1}{\tau}|G_E|^2\sin^2\theta_{\bar\Lambda}  
 + \frac{Im(G_M G_E^*)}{\sqrt{\tau}}
   \sin(2\theta_{\bar\Lambda}) \left( S_\Lambda^y + S_{\bar\Lambda}^y\right)
 \nonumber \\
 &-   \frac{Re(G_M G_E^*)}{\sqrt{\tau}} 
   \sin(2\theta_{\bar\Lambda}) 
 \left( S_\Lambda^z S_{\bar\Lambda}^x + S_{\bar\Lambda}^z S_\Lambda^x\right)
 +  \left(\frac{1}{\tau}|G_E|^2+|G_M|^2\right)
   \sin^2\theta_{\bar\Lambda} \ \ S_{\bar\Lambda}^x  S_\Lambda^x
 \nonumber \\
 & +  \left(\frac{1}{\tau}|G_E|^2-|G_M|^2\right)
   \sin^2\theta_{\bar\Lambda} \ \ S_{\bar\Lambda}^y  S_\Lambda^y
-  \left(\frac{1}{\tau}|G_E|^2\sin^2\theta_{\bar\Lambda} 
  - |G_M|^2\left(1+\cos^2\theta_{\bar\Lambda}\right) \right) \ \
  S_{\bar\Lambda}^z  S_\Lambda^z
  \Biggr\} \ , 
\label{tensorsrestframes} 
  \eea
\end{widetext}
where $S_\Lambda^i$ ($S_{\bar\Lambda}^i$) is the $i$th component of the unit
 vector pointing into the direction of the $\Lambda$ ($\bar\Lambda$) spin,
 calculated in the  $\Lambda$ ($\bar\Lambda$) rest frame, and 
 $\theta_{\bar\Lambda}$ is the $\bar\Lambda$ polar angle in the $e^+e^-$
 center of mass frame. The $y$-direction
 is perpendicular to the
 production plane with the positive direction defined through 
  $\bar e_y = \bar e_{e^+} \times \bar e_{\bar \Lambda}$, where
 $\bar e_{e^+}$ and $\bar e_{\bar \Lambda}$ are the unit vectors
 pointing into the positron and $\bar \Lambda$ momenta directions
 respectively (see Fig. 1).
\begin{figure}[hb]
 \vspace{0.5 cm}
\begin{center}
\includegraphics[width=8.5cm,height=6cm]{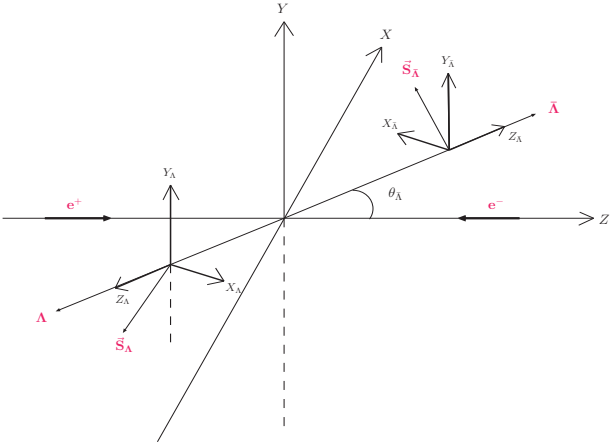}
\caption{(color online). Definition of the three reference frames used to
 obtain Eq. (\ref{tensorsrestframes}).
}
\end{center}
 \vspace{0.5 cm}
\label{f00}
\end{figure}

  The linear polarization is a consequence of the relative phase 
  between electric and magnetic form factors and points into the direction
 normal to the production plane \cite{Dubnickova:1992ii,Brodsky:2003gs}
 (and references therein). The correlation coefficients 
 $C_{x,z},C_{x,x}, C_{y,y}, C_{z,z}$ are in general non vanishing even if
 the relative phase between $G_E$ and $G_M$ is zero or $\pi$. 
 
  The absolute values of the electric and magnetic form factors 
 can be measured without measuring the polarization of the baryons
 via the Rosenbluth method, i.e. by analyzing 
 the angular distributions of the baryons. In previous investigations
 it was shown \cite{Dubnickova:1992ii} that the relative phase
 can be determined by using longitudinally polarized beams.
  From Eq.(\ref{tensorsrestframes}) it is obvious
  that a measurement of $S^y$, which can be performed for $\Lambda$, as
  discussed below, determines that phase modulo $\pi$. The remaining
 twofold ambiguity can be eliminated through a study of the correlation between
 $S_\Lambda^z$ and $S_{\bar\Lambda}^x$ 
 (or $S_\Lambda^x$ and $S_{\bar\Lambda}^z$ ), which is proportional
 to $Re(G_M G_E^*)$.

  The measurement of the subsequent
  two body decays of $\Lambda \to \pi^-p, \ \pi^0n$ and 
 its charge conjugate allow for a straightforward
 spin analysis of the decaying $\Lambda s$. The neutral modes
 are experimentally more demanding. In the remaining part of this work
  we will therefor refer only to the charged
 modes  even if the theoretical description
 of both is identical.
  The decay  distribution, when the spin of the proton is
  not measured, is proportional
  to 
\bea 
R_{\Lambda}
 =1-\alpha_{\Lambda}\ \bar S_{\Lambda}\cdot \bar n_{\pi^-}
 \ ,
\eea 
 where $\bar n_{\pi^-}$ is a unit vector in the direction of the 
 $\pi^-$  momentum. The factor $\alpha_{\Lambda}=0.642(13)$ characterizes
 the analyzing power of the angular distribution. For the antiparticle
 $\alpha_{\bar\Lambda}= -\alpha_{\Lambda}$.
 
 To combine production and decay, one sums over the intermediate
 spins of the $\Lambda$ and $\bar\Lambda$ and uses the fact that the 
 sums over the $\Lambda$ polarizations can be performed easily,
 \bea
  \sum_{pol.} \bar S_\Lambda =0 \ , \ \ \ \ \ \ 
  \sum_{pol.}  S^i_\Lambda S^j_\Lambda = \delta^{ij} \ .
 \eea
  The spin vectors in the Eq.(\ref{tensorsrestframes}) are then
  effectively replaced by  
  $\bar S_{\Lambda} \to -\alpha_\Lambda \bar n_{\pi^-}$ and
  $\bar S_{\bar\Lambda} \to -\alpha_{\bar\Lambda} \bar n_{\pi^+}$.

   As a result the relation between cross sections of the reaction
  $e^+e^-\to\bar\Lambda\Lambda$ and the reaction
  $e^+e^-\to\bar\Lambda(\to \pi^+\bar p)\Lambda(\to \pi^- p)$, 
 when the spins of the final proton and antiprotons are not measured, is
  simple.   Using the
narrow width approximation one finds
\bea
& \kern-40pt
d\sigma\left(e^+e^-\to\bar\Lambda(\to \pi^+\bar p)\Lambda(\to \pi^- p)\right)=
\nonumber \\ & \kern+35pt d\sigma\left(e^+e^-\to\bar\Lambda\Lambda\right)
 (S_{\Lambda,\bar\Lambda} \to \mp\alpha_\Lambda n_{\pi^\mp})
 \nonumber \\
& \times \ 
 d\bar\Phi_2(q_1;p_{\pi^+},p_{\bar p})d\bar\Phi_2(q_2;p_{\pi^-},p_{ p})
 \nonumber \\
& \kern-12pt \times \ {\rm Br}(\bar\Lambda \to \pi^+\bar p)
 {\rm Br}(\Lambda \to \pi^-  p) \ ,
\label{crsectrel}
\eea
where $d\bar\Phi_2$ is the two body phase space normalized to 1,
 $\alpha_\Lambda$  is the asymmetry parameter of the $\Lambda$ and
  $n_{\pi^+}$($n_{\pi^-}$) is a four vector, which in the $\bar\Lambda $ 
 ($\Lambda $) rest frame is equal to $(0,\bar  n_{\pi^+})$
 ($(0,\bar  n_{\pi^-})$), with
  $\bar n_{\pi^+} $
 ($\bar n_{\pi^-} $) being the unit vector in the direction
  of $\bar p_{\pi^+} $ ($\bar p_{\pi^-} $).
 The symbol $(S_{\Lambda,\bar\Lambda} \to \mp\alpha_\Lambda n_{\pi^\mp})$
 indicates that $S_{\Lambda}$ has to be replaced by 
 $-\alpha_\Lambda n_{\pi^-}$ and   $S_{\bar\Lambda}$ by 
 $\alpha_\Lambda n_{\pi^+}$.

 Let us now present a qualitative discussion of the observables
 for the radiative return.
 In \cite{Nowak} it was shown that in the case
 of  single photon emission from the initial state leptons, by choosing
 as reference frame the hadronic rest frame with the z-axis opposite
 to the photon direction 
 (we will call that frame RF from now on),
  one gets (for $Q^2\ll s$, which is fulfilled at 
 B--factories) the following simple approximate
 expression for the leptonic tensor
\begin{equation}
L^{ij}\simeq\frac{(4\pi\alpha)^2}{Q^4 y_1 y_2}
\frac{(1+\cos^2\theta_{\gamma})}{2} \mbox{\rm diag}(1,1,0) \ ,
\label{simplepttens}
\end{equation}
which is sufficient for a qualitative discussion.
The y-axis is chosen in the
 $e^+e^-$ cms frame as $\bar e_y = \bar e_{e^+} \times \bar e_{\gamma}$,
 where $\bar e_{e^+}$ and $\bar e_{\gamma}$ are  unit vectors
  pointing the directions of the positron and photon momenta, respectively.
 It remains unchanged after the transformation from the $e^+e^-$ cms frame
 to the hadronic rest frame described above.
Here $y_{1,2}=\frac{s-Q^2}{2s}(1\mp \cos\theta_{\gamma})$, 
$\theta_{\gamma}$ is the
 photon polar angle in the $e^+e^-$ center of mass frame
 and $i,j=1,2,3$ in $L^{ij}$.
 Other components of $L^{ij}$ do not contribute to the matrix element.
The close similarity with Eq. (\ref{l0}) is obvious.
As a result 
\begin{widetext}
\bea
  L^{ij}H_{ij} \simeq &\frac{(4\pi\alpha)^3}{4 Q^2 y_1 y_2}
 \left(1+\cos^2\theta_{\gamma}\right) 
\Biggl\{
 |G_M|^2\left(1+\cos^2\theta_{\bar\Lambda}\right)
 +\frac{1}{\tau}|G_E|^2\sin^2\theta_{\bar\Lambda}  
- \alpha_\Lambda\frac{Im(G_M G_E^*)}{\sqrt{\tau}}
   \sin(2\theta_{\bar\Lambda}) \left( n_{\pi^-}^y - n_{\pi^+}^y\right)
 \nonumber \\
 &+ \alpha_\Lambda^2  \frac{Re(G_M G_E^*)}{\sqrt{\tau}} 
   \sin(2\theta_{\bar\Lambda}) 
 \left( n_{\pi^-}^z n_{\pi^+}^x + n_{\pi^+}^z n_{\pi^-}^x\right)
 - \alpha_\Lambda^2 \left(\frac{1}{\tau}|G_E|^2+|G_M|^2\right)
   \sin^2\theta_{\bar\Lambda} \ \ n_{\pi^+}^x  n_{\pi^-}^x
 \nonumber \\
 & - \alpha_\Lambda^2 \left(\frac{1}{\tau}|G_E|^2-|G_M|^2\right)
   \sin^2\theta_{\bar\Lambda} \ \ n_{\pi^+}^y  n_{\pi^-}^y
 + \alpha_\Lambda^2 \left(\frac{1}{\tau}|G_E|^2\sin^2\theta_{\bar\Lambda} 
  - |G_M|^2\left(1+\cos^2\theta_{\bar\Lambda}\right) \right) \ \
  n_{\pi^+}^z  n_{\pi^-}^z
  \Biggr\} \ ,
\label{spin}
\eea
\end{widetext}
 with the angle $\theta_{\bar \Lambda}$
 being defined now in RF frame and  $n_{\pi^+}$ ($n_{\pi^-}$)
 in the $\bar \Lambda$ ($\Lambda$) rest frame.
  Hence after
 measuring $|G_M|$ and $|G_E|$ with the method
 proposed in \cite{Nowak} and successfully used by BaBar for the
 proton form factors \cite{Aubert:2005cb} 
 (no information on the spin is necessary in this case), one can measure
 the combination $\alpha_\Lambda \sin(\Delta\phi)$, where $\Delta\phi$ is
 the relative phase of the two form factors. This can be achieved for example
 by measuring the asymmetry
 \bea
 {\cal A}_y^\pm = \frac{d\sigma( a^\pm>0) - d\sigma( a^\pm<0) }
            {d\sigma( a^\pm>0) + d\sigma( a^\pm<0)} \ ,
\label{asym}
\eea
where
 $a^{+(-)} = \sin(2\theta_{\bar\Lambda})\ n_{\pi^+(\pi^-)}^y $.
 Since
 $\alpha_\Lambda$ is known with good precision \cite{Yao:2006px},
  $\sin(\Delta\phi)$, and thus $\Delta\phi$,  is determined up to 
 a twofold ambiguity, which can be 
  resolved by measuring the correlations between
   $n_{\pi^-}^z$ and  $n_{\pi^+}^x$. 
  In principle,
 by studying  the double spin correlations one can measure both
 the sign of  $\cos(\Delta\phi)$ and $\alpha_\Lambda$. 
 However, the expected number of events
 at B--factories amounts to a few hundred to a few thousand distributed
 in the energy range $4m_\Lambda^2 < Q^2 < 10 \ {\rm GeV}^2$, depending on the
 accumulated luminosity and the  angular cuts in the experimental
 analysis (see Section \ref{feasibility}).
  The measurement of $\alpha_\Lambda$ thus cannot be competitive
 to the existing measurements \cite{Yao:2006px}. 

\section{ The implementation 
 into the event generator PHOKHARA}

 To allow for quantitative studies
 we have implemented the process into the event generator PHOKHARA 
\cite{Rodrigo:2001jr,Rodrigo:2001kf,Kuhn:2002xg,Czyz:PH03,Nowak,Czyz:PH04,Czyz:2004nq,Czyz:2005as}.
 This implementation will be available
  within the version PHOKHARA6.0.
 We use the leading order matrix element only, since
  the expected cross section
 is small and correspondingly the expected statistical precision
 is low.
 The radiative corrections are expected to be of the
 order of a few percent \cite{Nowak}. 
   The generation of events is based on
  the method used in the generator TAUOLA \cite{Jadach:1990mz},
  performing first the generation with the unpolarized 
   cross section $d\sigma(e^+e^-\to\Lambda\bar\Lambda)$
 and generating in the second step the proper distributions 
 of the decay products by evaluating 
  the ratio of the cross sections with and without spin effects.
 Standard tests (see for example \cite{Czyz:PH03}), where the generation
 results were compared with existing analytic formulae, assure the technical
 accuracy of the implementation at the level of 10$^{-4}$.

  The cross section $\sigma(e^+e^- \to \bar\Lambda\Lambda)$ 
  was measured at one energy \cite{Bisello:1990rf}  only, and with 
  poor accuracy. No form factor has been
  extracted. Thus one has to relay on 
  theoretical assumptions.
  For the isospin and SU(3) structure of the electromagnetic
  form factors of the baryon octet we adopt the model ansatz
  suggested in reference \cite{Korner:1976hv}.
  There all form factors are given in terms of $\omega$-, $\phi$-
  and $\rho$-dominated contributions. Indeed, the $\rho$ piece
  does not contribute in the case of $\bar\Lambda\Lambda$ production.
  The adopted model predicts real 
  form factors $G_M$ and  $G_E$ and thus zero relative phase
 $\Delta\phi$. 
 The size of the relative phase between
  $G_M$ and  $G_E$ does not alter the cross section and thus till now
 no  experimental information is available on $\Delta\phi$. 
 To show the effects of a possibly large relative phase we add an ad hoc phase
 in the analysis presented in the next section.

\section{\label{feasibility} Feasibility study for B- factories}

\begin{figure}[ht]
 \vspace{0.5 cm}
\includegraphics[width=8.5cm,height=6cm]{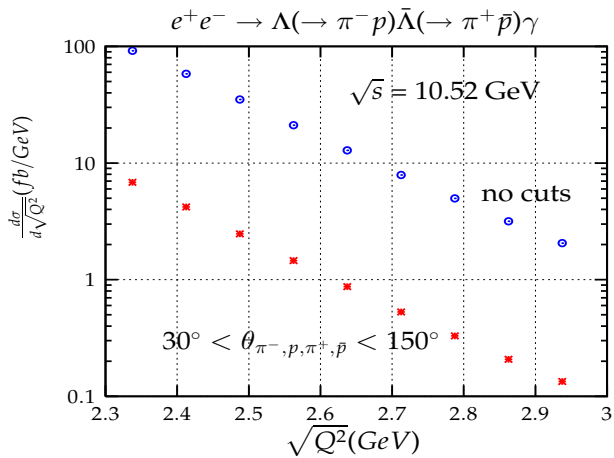}
\caption{(color online). The predicted differential cross section
 of the reaction 
 $e^+e^-\to \Lambda(\to \pi^-p) \bar \Lambda(\to \pi^+\bar p)\gamma$
}
 \vspace{0.5 cm}
\label{f1}
\end{figure}

 The predicted cross sections  as  functions
 of $Q^2$, with no cuts on the final state
  particles (open circles) and with angular cuts roughly corresponding
  to the angular range covered by the BaBar detector (filled circles),
  are shown in Fig. \ref{f1}. The integrated cross sections,
  for $\sqrt{Q^2}$ values from threshold up to 3 GeV, are equal to 18~fb
  when no angular cuts are applied and 1.3~fb, if we restrict the polar
 angles of the produced charged particles
  between 30 and 150 degrees. So typically at B-factories we
  expect about 130 events per 100~fb$^{-1}$ accumulated luminosity. 
  With the already accumulated luminosity at BaBar (390 fb$^{-1}$)
  and BELLE (710 fb$^{-1}$)
  the $\Lambda$ form factors can be studied with a decent accuracy.  
\begin{figure}[ht]
 \vspace{0.5 cm}
\includegraphics[width=8.5cm,height=6cm]{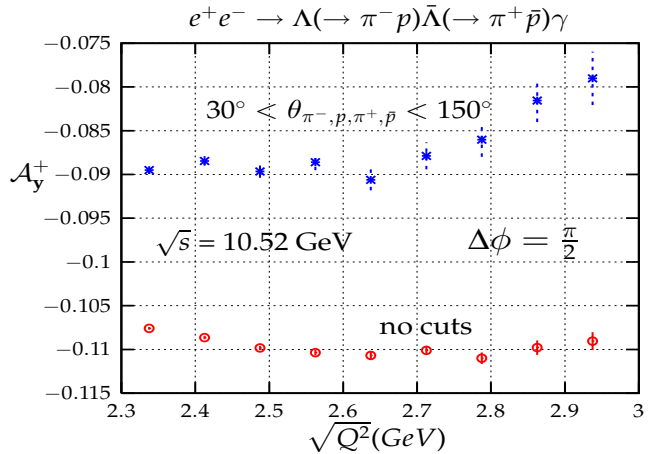}
\caption{(color online). 
 The asymmetry ${\cal A}_y^+$ (see Eq.(\ref{asym}) for definition)
 for the maximal relative phase ($\Delta\phi=\pi/2$)
  between the $G_M$ and  $G_E$
 form factors.
}
 \vspace{0.5 cm}
\label{f2}
\end{figure}

As already said,
 the adopted model predicts almost zero asymmetry ${\cal A}_y^\pm$.
 However,  a priori a much
 larger relative
 phase between the $G_M$ and  $G_E$ could be present. In Fig.\ref{f2}
 the predicted asymmetry is shown for the maximal relative phase
 between the form factors. It is clear that in this case the measurement
 of the phase is feasible. Thus the experimental studies of the  asymmetry
 will allow to obtain
  further constrains on the form factor models supplementing
 the information obtained by the $|G_M|$ and  $|G_E|$
 measurement through angular distributions of the $\Lambda$s.
 More refined strategies of this measurement are conceivable if the
 asymmetry is studied as a function of $\theta_\Lambda$. Furthermore,
 the sample is doubled by considering $\Lambda$ and $\bar\Lambda$
 decays.

The  combination $Re(G_M G_E^*)$ (or $\cos(\Delta\phi)$) can be extracted by
 measuring the asymmetry 

\bea
 {\cal A}_{xz} = \frac{d\sigma( \tilde a>0) 
 - d\sigma( \tilde a<0) }
            {d\sigma( \tilde a>0) + d\sigma( \tilde a<0)} \ , \nonumber \\
 \tilde a = \sin(2\theta_{\bar\Lambda}) 
 \left( n_{\pi^-}^z n_{\pi^+}^x + n_{\pi^+}^z n_{\pi^-}^x\right)\ .
\label{asymxz}
\eea

\begin{figure}[ht]
 \vspace{0.5 cm}
\includegraphics[width=8.5cm,height=6cm]{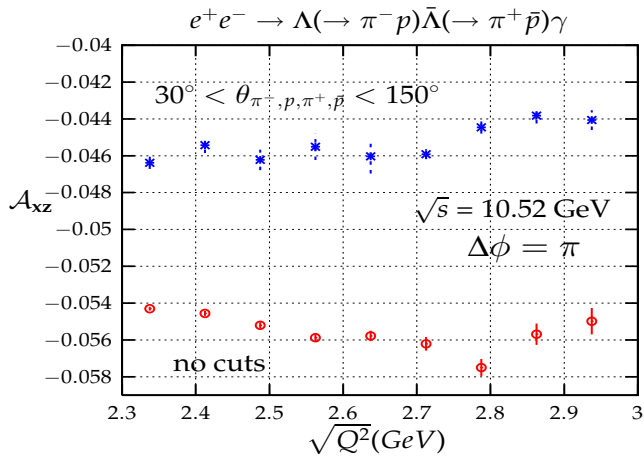}
\caption{(color online). 
 The asymmetry ${\cal A}_{xz}$ (see Eq.(\ref{asymxz}) for definition)
 for the maximal relative phase ($\Delta\phi=\pi$)
  between the $G_M$ and  $G_E$
 form factors.}
 \vspace{0.5 cm}
\label{f3}
\end{figure}

 Knowing both   $\sin(\Delta\phi)$ and $\cos(\Delta\phi)$ the relative phase
 is fixed. The predicted asymmetry $ {\cal A}_{xz}$ for $\Delta\phi=\pi$
 and a $B$- factory energy is shown in Fig. \ref{f3}.
 Measuring other spin correlations or respective asymmetries can 
 in principle allow for the measurement of $\alpha_\Lambda$
  and independent measurements of $|G_M|$ and  $|G_E|$. 
 On the one hand, 
 due to the limited statistics, correlations studies alone
 will not be competitive with the
 existing measurement of $\alpha_\Lambda$ \cite{Yao:2006px} or with the 
 extraction of $|G_M|$ and  $|G_E|$ by means of the method proposed
 in \cite{Nowak}.
 On the other hand, for each value of $Q^2$ the distribution,
 which is differential in $\cos\theta_{\Lambda}$ and in the variables
 describing the decay distributions of $\Lambda$ and $\bar \Lambda$,
 depends on the three quantities 
 $|G_M|$, $|G_E|$ and $\Delta\phi$ only. A more refined experimental
 investigation of this distribution, using the generator PHOKHARA,
 may well lead to a reasonably precise determination of these three numbers.

 \section{Summary}

 A compact formula has been presented for the spin dependent baryon 
 production through $e^+e^-$ annihilation  and through the radiative return.
 It describes the dependence on a single spin as well 
 as spin-spin correlations.
Single spin asymmetries can be used to determine the relative phase
 between the electric and the magnetic form factor up to a twofold 
 ambiguity, which can be resolved by analyzing spin--spin correlations.
 The construction of an extension of the Monte Carlo event generator PHOKHARA
 allowed us to demonstrate that the spin dependent decay 
 distribution
 of $\Lambda$ (and $\bar \Lambda$) can be used to perform this analysis
 under realistic experimental conditions.
 Given enough data, the study could be extended in a straightforward way
 to $\bar\Sigma\Sigma$, $\bar\Xi\Xi$ production with the subsequent decays
$\Sigma^+ \to n \pi^+,p\pi^0$, $\Sigma^- \to n \pi^-$,
  $\Xi^0\to\Lambda\pi^0$ and $\Xi^-\to\Lambda\pi^-$ as typical examples.

\begin{acknowledgments}
Henryk Czy\.z is grateful for the support and the kind hospitality 
of the Institut f{\"u}r Theoretische Teilchenphysik
 of the Karlsruhe University. 
\end{acknowledgments}

\bibliography{biblio}

\end{document}